\documentclass[fleqn,twoside]{article}
\usepackage{espcrc2}

\usepackage{graphicx}

\newcommand{\beq}{\begin{equation}}
\newcommand{\eeq}{\end{equation}}
\newcommand{\beqn}{\begin{eqnarray}}
\newcommand{\eeqn}{\end{eqnarray}}

\title{
\vspace{-9mm} \rightline{\small ITEP-LAT/2003-20} \vspace{-2mm}
\rightline{\small September, 2003}
Self-tuning of the P-vortices
    \thanks{Talk presented by A.V.K. at Lattice 2003, Tsukuba}
  }

\author{
    F.V. Gubarev\address[ITEP]{Institute of Theoretical and  Experimental
    Physics, B.~Cheremushkinskaya~25, Moscow, 117259, Russia}
\address[Kanazawa]{ITP, Kanazawa University, Kanazawa 920-1192, Japan},
A.V. Kovalenko\addressmark[ITEP],
       M.I. Polikarpov\addressmark[ITEP],
S.N. Syritsyn\addressmark[ITEP]
        and
V.I. Zakharov \address[MPI]{Max-Planck Institut f\"ur Physik, F\"ohringer Ring
6, 80805, M\"unchen, Germany}\thanks{Work is partially supported by grants RFBR
02-02-17308, \uppercase{RFBR 01-02-17456, DFG-RFBR 436 RUS 113/739/0,
INTAS-00-00111} and \uppercase{CRDF} award \uppercase{RPI-2364-MO}-02.}
}

\begin{document}
\begin{abstract}
We observe that on the currently available lattices the non-Abelian action
associated with the P-vortices is ultraviolet divergent. On the other
hand, the total area of the vortices scales in physical units.
Since both the ultraviolet and infrared scales are manifested and
there is no parameter to tune, the observed phenomenon can be called
self tuning.

\vspace{1pc}
\end{abstract}

\maketitle
\section{INTRODUCTION}
The Abelian monopoles and central vortices seem to be most promising
effective degrees of freedom to explain the confinement,
for review and references see, e.g., \cite{greensite}.
Straightforward interpretation of the lattice data is made
difficult, however, by the fact that monopoles and vortices are
defined in terms
of projected fields. One still can, upon  detection of the vortices
or of monopoles by means of projections,
 measure their entropy and  non-Abelian action.
In case of the monopoles, it turns out that there is a non-trivial
behavior of the action in the ultraviolet, see \cite{anatomy} and
references therein. In other words, the monopoles can be observed
only as a result of cancellation between the mass and entropy,
for discussion see \cite{vz}. The point-like
facet of the monopoles is manifested also in properties
of the monopole clusters \cite{clusters}.

In this note we summarize the first steps made to realize a similar
program of studying the anatomy
of the P-vortices \cite{kovalenko}. Our main result is that the
non-Abelian action of the P-vortices is comparable to the action of
the zero-point fluctuations and is ultraviolet divergent in the
limit $a\to 0$:
\beq\label{result}
S_{vortex}~\approx~0.54\cdot A_{vort}\cdot a^{-2}~,
\end{equation}
where $A_{vort}$ is the area of the vortex, the measurements refer to the
$SU(2)$ case and the P-vortices defined in the so called indirect
central gauge, for details see \cite{kovalenko}. Note that naively
one would assume that for non-perturbative fluctuations
$S_{n-pert}~\sim ~A\cdot\Lambda_{QCD}^2$.

In Sect. 2 we present the results of the measurements while in
Sect. 3 we discuss in more detail structure of the P-vortices and
some implications of our measurements.

\section{RESULTS}
 We have performed our calculations in pure $SU(2)$ lattice
gauge theory for $2.35 \leq \beta\leq 2.6$. The lattice spacing $a$ is fixed
using $\sqrt{\sigma}~=~440MeV$ where $\sigma$ is the string tension. At each
value of $\beta$ we have considered 20 statistically independent configurations
generated on symmetric $L^4$ lattices. The lattice size was $L$=16 for
$\beta$=2.35, $L$=24 for $\beta$= 2.4,2.45,2.5 and $L$=28 at $\beta$=2.55,2.6.
The indirect maximal center gauge \cite{deldebbio} was employed to define the
P-vortices. For further details see \cite{kovalenko}.

In Fig. 1 we summarize the results of the measurements of the P-vortex
density, $\rho_{vort}\equiv\langle
N_{PV}/(6L^4a^2)\rangle$, where $N_{PV}$ is the number
of plaquettes occupied by the P-vortex. Alternatively,
$A_{vort}=6\rho_{vort}\cdot V_4$
where $V_4$ is the volume of the lattice. We see that the vortex
density
is approximately a constant in physical units:
\beq
\rho_{vort}~\approx~4~(fm)^{-2}~~.
\eeq
Note that the scaling of the vortex area was observed earlier
\cite{olejnik}. We confirm this result on better statistics and
for smaller values of $a$.

\begin{figure}
\begin{center}
\includegraphics[width=\columnwidth]{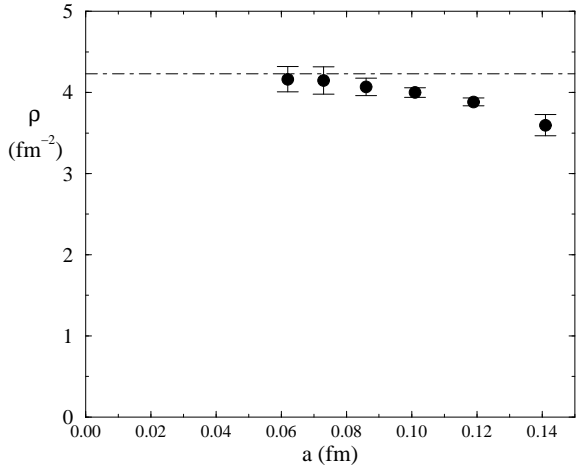}
\vspace{-12mm}
\caption {The density of P-vortices vs. lattice spacing
\vspace{-7mm}
}
\label{rho_fig}
\end{center}
\end{figure}

In Fig. 2 we summarize measurements of the non-Abelian action associated with
P-vortices. In more detail, we measure the average action density, $S_{PV}$, on
the plaquettes dual to those forming P-vortices. We subtract from it the
average plaquette action, $S_{vac}=\beta(1-\langle Tr~U_P\rangle /2)$. Both
$S_{PV}$ and $S_{vac}$ are measured in the lattice units. The difference
$(S_{PV}-S_{vac})=S_{vortex}$ is shown on the Fig.2 by circles and is
approximately a constant in the lattice units. Note that for particular value
of $\beta=2.4$ this difference was measured first in Ref. \cite{giedt} and we
agree with the results obtained therein.

\begin{figure}
\begin{center}
\includegraphics[width=\columnwidth]{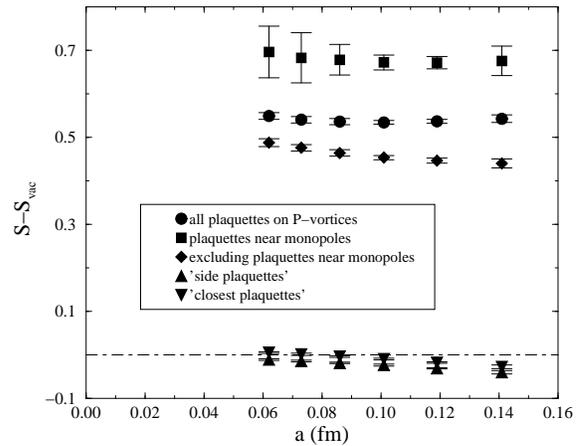}
\vspace{-12mm}
\caption {The excess of the non-Abelian action on and next to the P-vortices
\vspace{-7mm}
}
\label{action_fig}
\end{center}
\end{figure}

To further probe the structure of the vortices we have measured also
the average action density next to the P-vortex world
sheet. Geometrically, there are in fact two various types of
plaquettes neighboring the plaquettes belonging to the P-vortex,
for details see \cite{kovalenko}. We have measured the average action
for these two types of plaquettes separately (up and down triangles
on the Fig. 2). In the both cases,
the average action is very close to that of vacuum. Thus, we conclude
that the vortex thickness is smaller than the resolution available:
\beq
R_{vort}~\leq~0.06~fm~~,
\eeq
where $0.06~fm$ is the smallest lattice spacing used in our
simulations.

\section{SELF TUNING}
Thus, the vortices represent quite a unique type of the vacuum
fluctuations which exhibits both the ultraviolet and infrared scales.
Let us remind the reader that the most common examples demonstrate
sensitivity to one of the two scales. Thus, zero-point fluctuations
are controlled by the ultraviolet scale $1/a$. In case of instantons
both the effective size and action are determined by $\Lambda_{QCD}$.
Much less common are examples of fine tuned fluctuations when
one can tune a parameter to ensure that, say, inverse size is much
larger than the mass. For example, for the monopole mass in case
of the compact $U(1)$ one has
\beq
m^2_{mon}~\approx~const (e^{-2}-e^{-2}_{crit})a^{-2}~~,
\eeq
where $e$ is the electric charge and $e^2_{crit}$ is fixed
(for a review and references see, e.g., \cite{vz}). Tuning $e^2$
to $e^2_{crit}$ one ensures that $m^2_{mon}\ll a^{-2}$.

In case of a non-Abelian theory there is no parameter to tune since the
coupling is running. Nevertheless, the radiative mass of the monopole is
largely cancelled by the entropy, for discussion see \cite{clusters,vz}. Thus,
the underlying gauge theory ensures a kind of self tuning for the monopoles.

Now we observe a similar phenomenon for the P-vortices: their area
scales in the physical units while the tension is a constant in the
lattice units. Suppression of the vortices due to their action is
approximately
\beq\label{aaction}
\exp(-S_{vortex})~\approx~\exp(~-0.54\cdot A/a^2)~~.
\eeq
For the area of the vortices not to shrink this factor is to be cancelled by
the entropy. P-vortices represent closed surfaces and their entropy can be
estimated knowing the position of the phase transition to the vortex
percolation in case of the lattice $Z(2)$:
\beq\label{entropy}
(Entropy)_{vort}~\approx~\exp(~+0.88\cdot A/a^2)~~,
\eeq
(for discussion of the $Z(2)$ theories and further references
see, e.g., \cite{creutz}).

Apparently, the action and entropy factors do not cancel each other to the
accuracy needed. In other words, the self tuned vortices cannot be described in
terms of a local action density alone but are to possess further structure.
And, indeed, it was observed, for a particular $\beta$ that the vortices are
populated by the monopoles \cite{giedt}. We confirm this picture for all the
values of $\beta$ tested. Moreover, the plaquettes shared by the monopoles and
vortices have even higher action than the P-vortices on average, see Fig. 2.
Accounting for the monopoles might bridge the gap between (\ref{entropy}) and
(\ref{aaction}) although no explicit calculation is available.

Our final remark is that we defined the vortex size in terms of the
distribution of the non-Abelian action. Another manifestation of the
small physical size of the vortices is the linearity of the heavy
quark potential associated with vortices \cite{linear}. However, one
could define thickness of the vortices it terms of their flux,
for discussion see \cite{deldebbio,greensite}. Then the vortex seems to be
not localized to the ultraviolet cut off. Indeed, the action which we
observe
is still considerably smaller than that of the negative plaquettes.

In conclusion let us mention that further observations on the interplay
between
the monopoles and vortices were presented in another talk
at this conference \cite{syritsyn}.


\begin{thebibliography}{99}
\bibitem{greensite}
J. Greensite, ``The confinement problem in lattice gauge theory'',
{\tt hep-lat/0301023}.

\bibitem{anatomy}
V.G.~Bornyakov {\rm et al.},  Phys. Lett.  B537 (2002) 291, {\tt
hep-lat/0103032}.

\bibitem{vz}
V.I. Zakharov, ``Hidden mass hierarchy in QCD'', {\tt hep-ph/020204};
``Branes in lattice SU(2) gluodynamics'', {\tt hep-ph/0306261}.


\bibitem{clusters}
M.N. Chernodub and V.I. Zakharov, Nucl. Phys. B669 (2003) 233,
{\tt hep-th/0211267};
V.G. Bornyakov, P.Yu. Boyko, M.I. Polikarpov and V.I. Zakharov,
''Monopole clusters at short and large distances'',  {\tt hep-lat/0305021}.

\bibitem{kovalenko}
F.V. Gubarev, A. V. Kovalenko, M. I. Polikarpov, S. N. Syritsyn and
V. I. Zakharov,``Fine tuned vortices in lattice $SU(2)$
gluodynamics'', {\tt hep-lat/0212003}.


\bibitem{deldebbio}
L. Del Debbio, M. Faber, J. Greensite and S. Olejnik, Phys. Rev. D55
(1997) 2298, {\tt hep-lat/9610005}.

\bibitem{olejnik}
L. Del Debbio {\rm et al.}, Phys. Rev. D58 (1998) 094501,
{\tt hep-lat/9801027}.

\bibitem{giedt}
J. Ambjorn, J. Giedt and J. Greensite, JHEP 0002 (2000) 033,
{\tt hep-lat/9907021}.

\bibitem{creutz}
M. Creutz, L. Jacobs and C.V. Rebbi, Phys. Rev. Lett., 42 (1979) 1390.

\bibitem{linear}
F.V. Gubarev, M.I. Polikarpov and V.I. Zakharov, Mod. Phys. Lett A14
(1999) 2039, {\tt hep-th/9812030};
M.N. Chernodub, F.V. Gubarev, M.I. Polikarpov and V.I. Zakharov, Phys.
Lett. B475 (2000) 303, {\tt hep-ph/0003006}.

\bibitem{syritsyn}
A. V. Kovalenko, M. I. Polikarpov, S. N. Syritsyn and V. I. Zakharov,
``Interplay of Monopoles and P-Vortices'', preprint ITEP-LAT/2003-23;
proceedings of this conference.

\end{thebibliography}
\end{document}